\documentclass[11pt,a4paper]{article}
\usepackage{jstyle}

\newcommand{\bi}{\begin{itemize}}
\newcommand{\ei}{\end{itemize}}
\newcommand{\bea}{\begin{align}}
\newcommand{\eea}{\end{align}}
\newcommand{\be}{\begin{equation}}
\newcommand{\ee}{\end{equation}}

\newcommand{\pl}{{\partial}}

\makeatletter
\renewcommand*\env@matrix[1][\arraystretch]{%
  \edef\arraystretch{#1}%
  \hskip -\arraycolsep
  \let\@ifnextchar\new@ifnextchar
  \array{*\c@MaxMatrixCols c}}
\makeatother

\author{Charlotte SLEIGHT$^{a}$}

\author{\quad Massimo TARONNA$^{b}$ \footnote{Postdoctoral Researcher of the Fund for Scientific Research-FNRS Belgium.}}

\affiliation{${}^{a}$Max-Planck-Institut f\"ur Physik\\
\hspace*{0.01cm} F\"ohringer Ring 6, 80805 Munich, Germany}

\affiliation{${}^{b}$Universit\'e Libre de Bruxelles
and International Solvay Institutes\\
\hspace*{0.01cm} ULB-Campus Plaine CP231, 1050 Brussels, Belgium}

\emailAdd{charlotte.sleight@gmail.com, massimo.taronna@ulb.ac.be}

\title{\centering
\LARGE{Higher-spin Interactions from CFT:}\\
\LARGE{The Complete Cubic Couplings}}

\abstract{In this letter we provide a complete holographic reconstruction of the cubic couplings in the minimal bosonic higher-spin theory in AdS$_{d+1}$. For this purpose we also determine the OPE coefficients of all single-trace conserved currents in the $d$-dimensional free scalar $O\left(N\right)$ vector model, and compute the tree-level three-point Witten diagram amplitudes for a generic cubic interaction of higher-spin gauge fields in the metric-like formulation.}

\begin{document}
\begin{flushright}    
  {MPP-2016-25}
\end{flushright}
\maketitle

\section{Introduction}
Despite the long existence of fully non-linear equations of motion \cite{Vasiliev:1990en} for theories of higher-spin gauge fields, a complete list of Lagrangian cubic couplings is unknown. The main difficulty has been in taking the Noether procedure beyond the cubic order. Indeed, cubic consistency leaves the relative coefficients of the possible cubic interactions arbitrary. So far only a small number of Lagrangian couplings have been determined, in particular those which are constrained by the higher-spin algebra structure constants \cite{Fradkin:1986qy,Vasilev:2011xf,Boulanger:2013zza,Kessel:2015kna,hsalg}.\\

\noindent
In this letter we take an alternative route, and find a complete list of cubic couplings for the type A minimal bosonic higher-spin theory in anti-de Sitter (AdS) space as determined by holographic reconstruction. For higher-spin symmetry preserving boundary conditions, the latter theory is conjectured \cite{Sezgin:2002rt,Klebanov:2002ja} to be dual to (the singlet sector of) the free scalar $O\left(N\right)$ vector model. For the duality to hold, the cubic interactions of the bulk theory must reproduce the three-point correlators of the dual CFT. From this basic requirement, we are able to determine all bulk cubic couplings of the minimal bosonic higher-spin theory. As a non-trivial check of our holographic reconstruction, we relate the couplings obtained to the metric-like classification of gauge invariant bulk couplings in \cite{Joung:2011ww,Joung:2012fv,Joung:2013nma}.\\

\noindent
To reach our goal of a complete set of cubic couplings, we established two main intermediate results:
\begin{enumerate}
    \item The explicit form of all two- and three-point functions of conserved currents in the singlet sector of the $d$-dimensional free scalar $O\left(N\right)$ vector model.
    \item The amplitude of a three-point Witten diagram for an arbitrary triplet $\left(s_1,s_2,s_3\right)$ of external higher-spin gauge fields, generated by a generic consistent cubic interaction (with arbitrary relative coefficients).
\end{enumerate}
 
\noindent
This work constitutes a key step of  the  holographic reconstruction program, which began with the extraction of the $0$-$0$-$s$ cubic couplings \cite{Petkou:2003zz,Bekaert:2015tva,Skvortsov:2015lja,Taronna:2016ats} and scalar quartic self-interaction \cite{Bekaert:2014cea,Bekaert:2015tva,Bekaert:2016ezc}. See also the subsequent work \cite{Skvortsov:2015pea}, which gave a simple extension of the cubic $0$-$0$-$s$ results to the type B theory.\footnote{In 4d, the type B theory contains a parity odd scalar, as opposed to the parity even scalar of the type A theory under consideration. The type B theory is conjectured to be dual to the Gross-Neveu model \cite{Leigh:2003gk,Sezgin:2003pt}.} We give a more detailed summary of the present findings in the following section.

\section*{Overview and summary of results}

\subsection*{Complete conserved current three-point functions}

In CFT one can parametrise the most general conformal structure built from three distinct points in terms of the following six basic objects
{\allowdisplaybreaks
\begin{subequations}
\label{6conf}
\begin{align} 
{\sf Y}_1&=\frac{z_1\cdot x_{12}}{x_{12}^2}-\frac{z_1\cdot x_{13}}{x_{13}^2}\,,& {\sf H}_1&=\frac{1}{x_{23}^2}\left(z_2\cdot z_3+\frac{2 z_2\cdot x_{23}\,z_3\cdot x_{32}}{x_{23}^2}\right)\,,\\
{\sf Y}_2&=\frac{z_2\cdot x_{23}}{x_{23}^2}-\frac{z_2\cdot x_{21}}{x_{21}^2}\,,&{\sf H}_2&=\frac{1}{x_{31}^2}\left(z_3\cdot z_1+\frac{2 z_3\cdot x_{31}\,z_1\cdot x_{13}}{x_{31}^2}\right)\,,\\
{\sf Y}_3&=\frac{z_3\cdot x_{31}}{x_{31}^2}-\frac{z_3\cdot x_{32}}{x_{32}^2}\,,& {\sf H}_3&=\frac{1}{x_{12}^2}\left(z_1\cdot z_2+\frac{2 z_1\cdot x_{12}\,z_2\cdot x_{21}}{x_{12}^2}\right)\,.
\end{align}
\end{subequations}}

\noindent
By demanding the correct spin structure and behaviour under scale transformations, conformal symmetry dictates that the three-point function of a generic triplet of conserved currents is a sum of monomials of the type
\begin{multline}
\langle {\cal J}_{s_1}(x_1|z_1){\cal J}_{s_2}(x_2|z_2){\cal J}_{s_3}(x_3|z_3)\rangle\\=\sum_{n_i}{\sf C}_{s_1,s_2,s_3}^{n_1,n_2,n_3}\frac{{\sf Y}_1^{s_1-n_2-n_3}{\sf Y}_2^{s_2-n_3-n_1}{\sf Y}_3^{s_3-n_1-n_2}{\sf H}_1^{n_1}{\sf H}_2^{n_2}{\sf H}_3^{n_3}}{(x_{12}^2)^{\tfrac{\tau_1+\tau_2-\tau_3}{2}}(x_{23}^2)^{\tfrac{\tau_2+\tau_3-\tau_1}{2}}(x_{31}^2)^{\tfrac{\tau_3+\tau_1-\tau_2}{2}}}\,,
\end{multline}
with twists $\tau_i=\Delta_i-s_i$ and theory-dependent coefficients ${\sf C}^{n_1,n_2,n_3}_{s_1,s_2,s_3}$.
\\

\noindent
We find that the explicit form of the three-point functions involving any triplet of conserved currents \eqref{jf} in the $d$-dimensional free scalar $O\left(N\right)$ vector model can be expressed in the following factorised form
\begin{multline}\label{dual3pt}
\langle {\cal J}_{s_1}(x_1|z_1){\cal J}_{s_2}(x_2|z_2){\cal J}_{s_3}(x_3|z_3)\rangle\\=N \left(\prod_{i=1}^3{\sf c}_{s_i}\,q_i^{\frac{1}{2}-\frac{\Delta }{4}}\,\Gamma(\tfrac{\Delta}{2})\,
J_{\frac{\Delta -2}{2}}\left(\sqrt{q_i}\right) \right)
\,\frac{{\sf Y}_1^{s_1}{\sf Y}_2^{s_2}{\sf Y}_3^{s_3}}{(x_{12}^2)^{\Delta/2}(x_{23}^2)^{\Delta/2}(x_{31}^2)^{\Delta/2}}\,,
\end{multline}
with  
\begin{align}
q_1&=2{\sf H}_1\pl_{{\sf Y}_2}\pl_{{\sf Y}_3}\,,&q_2&=2{\sf H}_2\pl_{{\sf Y}_3}\pl_{{\sf Y}_1}\,,&q_3&=2{\sf H}_3\pl_{{\sf Y}_1}\pl_{{\sf Y}_2}\,.\label{x}
\end{align}
With the choice of unit normalisation of the conserved current two-point functions \eqref{js2pt}, the OPE coefficients are the coefficients ${\sf C}_{s_1,s_2,s_3}^{(n_1,n_2,n_3)}$ of each conformal structure in \eqref{dual3pt}.
This fixes
\begin{align}
{\sf C}_{s_1,s_2,s_3}^{0,0,0}&=N\, \prod_{i=1}^3\, {\sf c}_{s_i}\,,& {\sf c}_{s_i}^2&= \frac{\sqrt{\pi}\,2^{-\Delta-s_i+3}\,\Gamma(s_i+\tfrac{\Delta}{2})\Gamma(s_i+\Delta-1)}{N\,s_i!\,\Gamma(s_i+\tfrac{\Delta-1}{2})\Gamma(\tfrac{\Delta}2)^2}\,,\label{g}
\end{align}
and the remaining coefficients are generated by the action of the derivative structures \eqref{x}

{\small
\begin{align}
{\sf C}_{s_1,s_2,s_3}^{n_1,n_2,n_3}=\frac{2^{-(n_1+n_2+n_3)}s_1!s_2!s_3!}{(s_1-n_2-n_3)!(s_2-n_3-n_1)!(s_3-n_1-n_2)!n_1!n_2!n_3!}\frac{{\sf C}_{s_1,s_2,s_3}^{0,0,0}}{(\tfrac{\Delta}{2})_{n_1}(\tfrac{\Delta}{2})_{n_2}(\tfrac{\Delta}{2})_{n_3}}. 
\end{align}}

\noindent
Remarks:
\begin{itemize}
    \item The above result is nicely factorised, as expected from a free CFT. Notice that the overall   coefficients \eqref{g} are simply a product of $s$-$0$-$0$ OPE coefficients. This generalises the result \cite{Diaz:2006nm} for the latter to three-point functions of an arbitrary triplet $\left(s_1,s_2,s_3\right)$ of single-trace operators.
    \item The Bessel function form \eqref{dual3pt} of tensor structures generated are consistent with those observed in the literature  for three-point functions of arbitrary spin conserved currents \cite{Osborn:1993cr,Erdmenger:1996yc,Stanev:2012nq,Zhiboedov:2012bm}.
    \item With the simple factorised result \eqref{dual3pt}, it is tempting to conjecture that the form of any $n$-point function of such currents is given by simply extending the product from $i=1,\ldots,n$
\begin{multline}
\langle {\cal J}_{s_1}(x_1|z_1){\cal J}_{s_2}(x_2|z_2)\ldots {\cal J}_{s_n}(x_n|z_n)\rangle= \frac{N}{(x_{12}^2)^{\Delta/2}(x_{23}^2)^{\Delta/2}\ldots (x_{n1}^2)^{\Delta/2}}\\\times\,\left(\prod_{i=1}^n{\sf c}_{s_i}\,q_i^{\frac{1}{2}-\frac{\Delta }{4}}
\Gamma(\tfrac{\Delta}{2})\,J_{\frac{\Delta -2}{2}}\left(\sqrt{q_i}\right) \right)
\,{\sf Y}_1^{s_1}{\sf Y}_2^{s_2}\ldots {\sf Y}_n^{s_n}+\text{perm.}\,,
\end{multline}
summing over all inequivalent permutations of the external legs.
\end{itemize}

\subsection*{Complete cubic couplings for minimal higher-spin theory}
The most general cubic vertex in AdS$_{d+1}$ is parameterised by six basic contractions. These are the bulk counterparts of the six conformal structures on the boundary \eqref{6conf}. Using point-splitting, the most general (on-shell) cubic interaction involving fields of spins $s_1$-$s_2$-$s_3$ is given by \cite{Joung:2011ww}:
\begin{multline}
    I_{s_1,s_2,s_3}^{n_1,n_2,n_3}(\Phi_i)=\mathcal{Y}_1^{s_1-n_2-n_3}\mathcal{Y}_2^{s_2-n_3-n_1}\mathcal{Y}_3^{s_3-n_1-n_2}\\\times\mathcal{H}_1^{n_1}\mathcal{H}_2^{n_2}\mathcal{H}_3^{n_3}\,\Phi_1(X_1,U_1)\Phi_2(X_2,U_2)\Phi_3(X_3,U_3)\Big|_{X_i=X}\,,
\end{multline}
with
\begin{subequations}
\begin{align}
    \mathcal{Y}_1&=\pl_{U_1}\cdot\pl_{X_2}\,,&\mathcal{Y}_2&=\pl_{U_2}\cdot\pl_{X_3}\,,&\mathcal{Y}_3&=\pl_{U_3}\cdot\pl_{X_1}\,,\\
    \mathcal{H}_1&=\pl_{U_2}\cdot\pl_{U_3}\,,&\mathcal{H}_2&=\pl_{U_3}\cdot\pl_{U_1}\,,&\mathcal{H}_3&=\pl_{U_1}\cdot\pl_{U_2}\,.
\end{align}
\end{subequations}
We work using the ambient space formulation of symmetric and traceless fields, which we briefly review in Section \ref{bulk} and Appendix \ref{radialred}. In appendix \ref{Appendix:deDonder} we complete the vertices obtained to the de Donder gauge. The most general such cubic interaction thus takes the form 
\begin{equation}\label{v123}
    {\cal V}_{s_1,s_2,s_3} = \sum\limits_{n_i} g^{n_1,n_2,n_3}_{s_1,s_2,s_3}I_{s_1,s_2,s_3}^{n_1,n_2,n_3}(\Phi_i).
\end{equation}
Above the space-time derivatives are flat ambient derivatives, and provide a simple choice for a basis of bulk structures (see e.g. \cite{Joung:2013nma}).\\

\noindent
Under the conjectured duality with the free scalar $O\left(N\right)$ vector model \cite{Klebanov:2002ja}, the tree-level Witten diagram generated by the bulk $s_1$-$s_2$-$s_3$ cubic vertex \eqref{v123} in type A minimal bosonic higher-spin theory must be equal to the dual three-point function \eqref{dual3pt} in the free scalar $O\left(N\right)$ vector model. Employing this correspondence, we determine the complete cubic action holographically to be
\begin{equation}\label{vert}
    \mathcal{V}=\sum_{s_1,s_2,s_3} g_{s_1,s_2,s_3}\,I_{s_1,s_2,s_3}^{\,0,0,0}(\Phi_i)\,,
\end{equation}
with
\begin{align}
   g_{s_1,s_2,s_3}= \pi ^{\frac{d-3}{4}}2^{\tfrac{3 d-1+s_1+s_2+s_3}{2}}\frac{1}{\sqrt{N}\, \Gamma (d+s_1+s_2+s_3-3)}\prod_{i=1}^3\sqrt{\frac{\Gamma(s_i+\tfrac{d-1}{2})}{\Gamma\left(s_i+1\right)}}\,,
\end{align}
and a canonical normalisation for the kinetic term of the higher-spin fields. Notice that in the chosen basis for bulk interactions the cubic vertex only involves structures with $n_i=0$. These terms however, once radially reduced, generate the necessary non-abelian quasi-minimal couplings of the HS theory (see e.g. \cite{Joung:2011ww,Joung:2013nma}). Nicely all these terms can be resummed into a single ambient structure (see Appendix \ref{radialred} for the reduction to intrinsic quantities and Appendix \ref{GaugeInv} for the relation with previous classifications and Appendix \ref{Appendix:deDonder} for the off-shell completion to the de Donder gauge).

\section{Single-trace OPE coefficients}

\label{boundary}

\subsection{Generating function for conserved currents}

We first introduce some useful technology for dealing with conserved currents of arbitrary spin in CFT, and their correlation functions. \\

\noindent
The conserved currents we consider in this paper reside in the singlet sector of the free scalar $O\left(N\right)$ vector model, which are traceless bi-linears in the fundamental scalar $\phi^a$ \cite{Anselmi:1999bb}:
\begin{equation}
    {\cal J}_{\mu_1 ... \mu_s}   \: \sim \: \phi^a \partial_{\mu_1}...\partial_{\mu_s} \phi^a + ..., \quad a = 1, ..., N\,. \label{jschema}
\end{equation}
In the above, the $\ldots$ are further singlet bi-linear structures, which ensure conservation and tracelessness. Moreover, these currents are non-trivial only for even spins $s$.\\

\noindent
It is convenient to use index-free notation, introducing a null polarisation vector
\begin{equation}
    {\cal J}_s\left(x|z\right) \equiv {\cal J}_{\mu_1 ... \mu_s}z^{\mu_1} ... z^{\mu_s}\,, 
\end{equation}
where tracelessness is encoded in the null condition $z^2=0$. They can then be packaged in the compact expression,
\begin{equation}
    {\cal J}_s\left(x|z\right) = f^{(s)}\left(z \cdot \partial_{x_1},z \cdot \partial_{x_2}\right) \phi^a\left(x_1\right)\phi^a\left(x_2\right)\big|_{x_1,x_2 \rightarrow x}\,, \label{jf}
\end{equation}
where the function $f^{(s)}\left(x,y\right)$ is given in terms of a Gegenbauer polynomial,
\begin{equation}
    f^{(s)}\left(x,y\right) = \left(x+y\right)^s\,C^{(\tfrac{\Delta-1}{2})}_s\left(\frac{x-y}{x+y}\right)\,,
\end{equation}
and $\tfrac{\Delta}{2}$ denotes the scaling dimension of $\phi^a$. This is an old trick \cite{Craigie:1983fb}, and can easily be derived by demanding that the expression \eqref{jf} is annihilated by the conformal boost operator. This gives the differential equation for $f^{(s)}(x,y)$, 
\begin{equation}
\left[(\tfrac{\Delta}{2}+x\,\pl_x)\pl_x +(\tfrac{\Delta}{2}+y\,\pl_y)\pl_y\right]f^{(s)}(x,y)=0\,,
\end{equation}
whose solution is expressed in terms of the Gegenbauer polynomials above.\footnote{The above equation can be generalised to deal with scalar operators made of constituents with different dimensions $\Delta_1/2$ and $\Delta_2/2$ respectively. The corresponding primary is in this case a Jacobi polynomial $f(x,y)=(x+y)^sP_s^{(\Delta_2/2-1,\Delta_1/2-1)}(\tfrac{x-y}{x+y})$.}
For $\Delta=d-2$, which is twice the dimension of a free boson, the primary operator has dimension $d-2+s$ and saturates the unitarity bound for $s>0$. For this scaling dimension we thus obtain conserved currents, which is straightforward to verify.\\

\noindent
The form \eqref{jf} of the conserved currents plays a crucial role in the following sections, as it allows for the seamless application of Wick's theorem to determine their two- and three-point functions.

\iffalse
For this purpose the explicit form for the function $f^{(s)}(x,y)$ will also be useful
\begin{subequations}
\begin{align}
f^{(s)}(x,y)=\sum_{n=0}^s&c_{s,n}x^{s-n}y^n\,,\\
c_{s,n}=\frac{\sqrt{\pi}\,\Gamma(\tfrac{\Delta}{2}+s)\Gamma(\Delta+s-1)}{2^{\Delta-2}\Gamma(\tfrac{\Delta-1}{2})}&\frac{(-1)^n}{n!(s-n)!\Gamma(\frac{\Delta}{2}+n)\Gamma(s-n+\tfrac{\Delta}{2})}\,.
\end{align}
\end{subequations}
\fi

\subsection{Two-point functions}

Conformal symmetry fixes two-point functions up to an overall coefficient, with those of spin-$s$ conserved currents \eqref{jf} taking the form
\begin{equation}
    \langle {\cal J}_{s}\left(x_1|z_1\right) {\cal J}_{s}\left(x_2|z_2\right)  \rangle = {\sf C}_{{\cal J}_s} \frac{{\sf H}_3^s}{\left(x^2_{12}\right)^{\Delta}}\,. \label{js2pt}
\end{equation}
We determine the overall coefficient ${\sf C}_{{\cal J}_s}$ for the currents \eqref{jf} in the following.\\

\noindent 
Since the theory is free, we may simply apply Wick's theorem to express the two-point function in terms of that of the fundamental scalar. We have
\begin{multline}
   \langle {\cal J}_s\left(x_1|z_1\right){\cal J}_s\left(x_2|z_2\right) \rangle = f^{(s)}(z_1\cdot\partial_{y_1},z_1\cdot\partial_{y_2}) f^{(s)}(z_2\cdot\partial_{{\bar y}_1},z_2\cdot\partial_{{\bar y}_2}) \\
   \times \left[ \langle \phi^{a}(y_1)\phi^{a}(y_2)\rangle \langle \phi^{b}({\bar y}_1)\phi^{b}({\bar y}_2)\rangle  \,+\, \phi^{a}(y_2) \leftrightarrow  \phi^{b}({\bar y}_1) \,+\, \phi^{a}(y_2) \leftrightarrow  \phi^{b}({\bar y}_1) \right]\,. 
\end{multline}
To extract the overall two-point coefficient, by conformal invariance 
it is sufficient to restrict attention to terms with zero contractions of the null auxiliary vectors. We thus set to zero $z_1\cdot z_2$, and match with the corresponding term in \eqref{js2pt}. The computation is drastically simplified using the Schwinger-parametrised form
\begin{equation}
    \langle \phi^a\left(x_1\right)\phi^b\left(x_2\right) \rangle = \frac{\delta^{ab}}{\left(x^2_{12}\right)^{\Delta/2}} = \frac{\delta^{ab}}{\Gamma\left(\frac{\Delta}{2}\right)} \int^{\infty}_{0} \frac{dt}{t}\,t^{\frac{\Delta}{2}} e^{-t x^2_{12}}\,,
\end{equation}
which gives 
\begin{align}
    {\sf C}_{{\cal J}_s} & =\left[\tfrac{1+\left(-1\right)^s}{2}\right]N\,2^{s+1}\frac{(\Delta-1)_s (\Delta-1)_{2s}}{\Gamma\left(s+1\right)}\,,
\end{align}
as a consequence of the Gegenbauer orthogonality relation.

\subsection{Three-point functions and OPE coefficients}

We now turn to the three-point functions of the conserved currents, and employ the same approach as for the two-point functions in the previous section: A combination of Wick's theorem and Schwinger parameterisation. For previous results in three-dimensions, see \cite{Giombi:2011rz,Colombo:2012jx,Didenko:2012tv}.

\subsection*{Three-point functions of conserved currents}

As explained in the introduction, conformal symmetry dictates that the three-point function of a generic triplet of conserved currents has the form 
\begin{multline}\label{corr}
\langle {\cal J}_{s_1}(x_1|z_1){\cal J}_{s_2}(x_2|z_2){\cal J}_{s_3}(x_3|z_3)\rangle\\=\sum_{n_i}{\sf C}^{n_1,n_2,n_3}_{s_1,s_2,s_3}\frac{{\sf Y}_1^{s_1-n_2-n_3}{\sf Y}_2^{s_2-n_3-n_1}{\sf Y}_3^{s_3-n_1-n_2}{\sf H}_1^{n_1}{\sf H}_2^{n_2}{\sf H}_3^{n_3}}{(x_{12}^2)^{\tfrac{\tau_1+\tau_2-\tau_3}{2}}(x_{23}^2)^{\tfrac{\tau_2+\tau_3-\tau_1}{2}}(x_{31}^2)^{\tfrac{\tau_3+\tau_1-\tau_2}{2}}}\,,
\end{multline}
built from the six basic conformal structures \eqref{6conf}. Each individual term is independently invariant under conformal transformations, and thus conformal symmetry alone does not determine the coefficients ${\sf C}^{n_1,n_2,n_3}_{s_1,s_2,s_3}$. Current conservation gives further constraints, however in the following we simply determine their explicit form using Wick's theorem.% We later confirm that the result is consistent with the form required by conservation.
\\

\noindent 
As with the two-point functions in the previous section, we may set $z_i \cdot z_j =0$. This gives
\begin{align}
&\langle {\cal J}_{s_1}(x_1|z_1){\cal J}_{s_2}(x_2|z_2){\cal J}_{s_3}(x_3|z_3)\rangle  \\ \nonumber
    &  = \frac{8N}{\Gamma\left(\frac{\Delta}{2}\right)^3} \int^{\infty}_0 \left(\prod\limits^3_{i=1} \frac{dt_i}{t_i}
    t^{\frac{\Delta}{2}}_i\right) f^{(s_1)}\left(-2t_3 z_1 \cdot x_{12},-2t_2 z_1 \cdot x_{13}\right)f^{(s_2)}\left(-2t_3 z_2 \cdot x_{21},-2t_1 z_2 \cdot x_{23}\right) \\ \nonumber
    & \hspace*{6.5cm} \times f^{(s_3)}\left(-2t_1 z_3 \cdot x_{32},-2t_2 z_3 \cdot x_{31}\right) e^{-t_1x^2_{23}-t_2 x^2_{31}-t_3 x^2_{12}}\\ \nonumber
    & = \sum_{n_i=0}^{s_i} {\sf D}^{n_1,n_2,n_3}_{s_1,s_2,s_3} \frac{(z_1\cdot x_{12})^{s_1-n_1}(z_2\cdot x_{21})^{s_2-n_2}}{(x_{12}^2)^{\tfrac{\Delta}{2}+s_1+s_2-n_1-n_2}}\frac{(z_1\cdot x_{13})^{n_1}(z_3\cdot x_{31})^{n_3}}{(x_{31}^2)^{\tfrac{\Delta}{2}+n_1+n_3}}\frac{(z_2\cdot x_{23})^{n_2}(z_3\cdot x_{32})^{s_3-n_3}}{(x_{23}^2)^{\tfrac{\Delta}{2}+s_3+n_2-n_3}},
\end{align}
where

{\footnotesize
\begin{align}
    & {\sf D}^{n_1,n_2,n_3}_{s_1,s_2,s_3} = \frac{N (-1)^{n_1 +n_2 +n_3 }2^{3+s_1+s_2+s_3}}{s_1!s_2!s_3!}\binom{s_1}{n_1}\binom{s_2}{n_2}\binom{s_3}{n_3} (\Delta-1)_{s_1}(\Delta-1)_{s_2}(\Delta-1)_{s_3}\\
    &\times \frac{ \Gamma \left(s_1 +\frac{\Delta }{2}\right) \Gamma \left(s_2 +\frac{\Delta }{2}\right) \Gamma \left(s_3 +\frac{\Delta }{2}\right)}{\Gamma \left(n_1 +\frac{\Delta }{2}\right) \Gamma \left(n_2 +\frac{\Delta }{2}\right)\Gamma \left(n_3 +\frac{\Delta }{2}\right) }\frac{\Gamma \left(n_1 +n_3 +\frac{\Delta }{2}\right) \Gamma \left(n_2 -n_3 +s_3 +\frac{\Delta }{2}\right) \Gamma \left(-n_1 -n_2 +s_1 +s_2 +\frac{\Delta }{2}\right)}{\Gamma \left(-n_1 +s_1 +\frac{\Delta }{2}\right) \Gamma \left(-n_2 +s_2 +\frac{\Delta }{2}\right) \Gamma \left(-n_3 +s_3 +\frac{\Delta }{2}\right)}\,.\nonumber
\end{align}}

\noindent
By matching with the corresponding expansion of the general form for the correlator \eqref{corr}, the following recursion relation for the coefficients ${\sf C}^{n_1,n_2,n_3}_{s_1,s_2,s_3}$ can be established:
\begin{align}
{\sf C}_{s_1,s_2,s_3}^{n_1,n_2,n_3}&=2^{-n_1-n_2-n_3}\Big[{\sf D}_{s_1,s_2,s_3}^{n_2,s_2-n_3,s_3-n_1}\\
&-\sum_{i_1+i_2+i_3=1}^{n_1+n_2+n_3}(-1)^{i_1 +i_2 +i_3 } \binom{i_1 +i_2 -n_1 -n_2 +s_3 }{i_1 } \binom{i_1 +i_3 -n_1 -n_3 +s_2 }{i_3 }\nonumber\\
& \hspace{50pt}\times\binom{i_2 +i_3 -n_2 -n_3 +s_1 }{i_2 } 2^{-i_1 -i_2 -i_3 +n_1 +n_2 +n_3 } {\sf C}_{s_1,s_2,s_3}^{(n_1 -i_1 ,n_2 -i_2 ,n_3 -i_3 )}\Big]\,,\nonumber
\end{align}
where the summation assumes $i_i\leq n_i$. This has solution:
\begin{multline}
{\sf C}_{s_1,s_2,s_3}^{n_1,n_2,n_3}=-\frac{N(-1)^{n_1 +n_2 +n_3 } 2^{s_1 +s_2 +s_3-(n_1 +n_2 +n_3 ) +3}}{n_1 ! n_2 ! n_3 !(s_1-n_2-n_3)!(s_2-n_3-n_1)!(s_3-n_1-n_2)!}\\\times\frac{\Gamma \left(s_1 +\frac{\Delta }{2}\right)  \Gamma \left(s_2 +\frac{\Delta }{2}\right)  \Gamma \left(s_3 +\frac{\Delta }{2}\right)}{\Gamma \left(n_1 +\frac{\Delta }{2}\right) \Gamma \left(n_2 +\frac{\Delta }{2}\right) \Gamma \left(n_3 +\frac{\Delta }{2}\right)}\,(\Delta-1)_{s_1}(\Delta-1)_{s_2}(\Delta-1)_{s_3}\,,
\end{multline}
whose $n_i$ dependence can be re-summed in terms of a Bessel function, giving the following compact form for the correlation function:
\begin{multline}
\langle {\cal J}_{s_1}(x_1|z_1){\cal J}_{s_2}(x_2|z_2){\cal J}_{s_3}(x_3|z_3)\rangle\\=N \left(\prod_{i=1}^3{\sf c}_{s_i}\,q_i^{\frac{1}{2}-\frac{\Delta }{4}}\Gamma(\tfrac{\Delta}2)\,
J_{\frac{\Delta -2}{2}}\left(\sqrt{q_i}\right) \right)
\,\frac{{\sf Y}_1^{s_1}{\sf Y}_2^{s_2}{\sf Y}_3^{s_3}}{(x_{12}^2)^{\Delta/2}(x_{23}^2)^{\Delta/2}(x_{31}^2)^{\Delta/2}}\,.
\end{multline}
Here
\begin{align}
q_1&=2\,{\sf H}_1\pl_{{\sf Y}_2}\pl_{{\sf Y}_3}\,,&q_2&=2\,{\sf H}_2\pl_{{\sf Y}_3}\pl_{{\sf Y}_1}\,,&q_3&=2\,{\sf H}_3\pl_{{\sf Y}_1}\pl_{{\sf Y}_2}\,,
\end{align}
and ${\sf c}_{s_i}$ are given in \eqref{g} for the canonically normalised current two-point function. I.e. by redefining ${\cal J}_{s_i} \rightarrow 1/\sqrt{{\sf C}_{{\cal J}_{s_i}}} {\cal J}_{s_i}$. A nice check at this point is that the result for the $s$-$0$-$0$ correlator coincides with that already given in the literature \cite{Diaz:2006nm}.

\section{Three-point Witten diagrams}
\label{bulk}
We now turn to the bulk side of the story. The three-point correlation functions of conserved currents computed in the previous section are dual to three-point Witten diagrams generated by cubic interactions of the corresponding bulk gauge fields. In this section we compute the three-point amplitudes for a generic such cubic interaction, thus providing the dictionary between bulk cubic couplings and the boundary CFT correlators.

\subsection{Brief review of the ambient space formalism}
It is convenient to employ the ambient space formalism, in which AdS$_{d+1}$ space is realised as a hyperboloid in an ambient $\left(d+2\right)$-dimensional Minkowski space
\begin{align}
    X^2+1&=0\,, \qquad X^0>0\,. \label{adsamb}
\end{align}
In this section we give a brief overview of the relevant aspects of this framework, and direct the unfamiliar reader to e.g. \cite{Costa:2011mg,Penedones:2010ue,Paulos:2011ie,Taronna:2012gb,Costa:2014kfa} for further details.\\

\begin{figure}[t]
  \centering
  \includegraphics[scale=0.45]{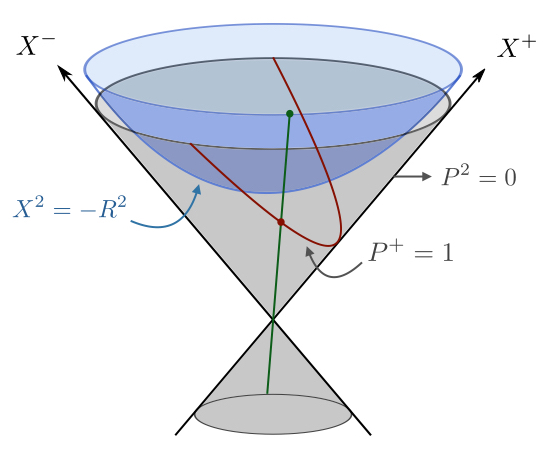} \label{amb}
  \caption{Euclidean AdS and its boundary in ambient space. This figure displays the
AdS surface $X^2 = - R^2=-1$ and the identification of a (green) boundary point with a (green) light ray
of the light cone $P^2
= 0$, which intersects the Poincar\'e section \eqref{ps} on a (red) point}
\end{figure}

\noindent
Fields intrinsic to AdS$_{d+1}$ space can be represented by homogeneous fields in ambient space that are tangent to the hyperboloid \eqref{adsamb}. As for the CFT discussion above, it is useful to encode tensor structures in polynomials of auxiliary variables. Symmetric rank-$s$ tensors can be described by 
\begin{equation}
    \Phi(X,U)=\frac1{s!}\,\Phi_{M_1\ldots M_s}(X)U^{M_1}\ldots U^{M_s}\,.
\end{equation}
subject to the following homogeneity and tangentiality conditions
\begin{align}
    (X\cdot\pl_X-\Delta)\Phi(X,U)&=0\,,& X\cdot\pl_U\Phi(X,U)&=0\,.
\end{align}
Above the degree of homogeneity for ambient symmetric fields is chosen to be compatible with the AdS/CFT dictionary for fields dual to conserved currents\footnote{Notice that this choice corresponds to the normalisable solution at the boundary of AdS$_{d+1}$. In \cite{Joung:2011ww} the opposite choice $\Delta=s-2$ was made, as it simplified the analysis of gauge invariance. The two choices are related by a change of basis in the space of couplings and are completely equivalent.}
\begin{equation}
    \Delta=2-d-s\,.
\end{equation}
The boundary of AdS$_{d+1}$ is described by the hypercone $P^2=0$. It is then convenient to introduce auxiliary variables $Z_A(x)$ to be contracted with the CFT currents defined on the hypercone at the boundary point $x$. The explicit relation between the ambient and intrinsic variables can be obtained solving the constraints in some given coordinate system. It is usually given employing ambient light cone coordinates $X^A=(X^+,X^-,X^a)$ as
\begin{align}
    Z^B(x)&=(0,2x\cdot z,z^b)\,,& P^B(x)&=(1,x^2,x^b)\,, \label{ps}
\end{align}
where $X^2=-X^+ X^- +\delta_{ab}X^a X^b$. We include figure \ref{amb} for clarity.

\subsection{Cubic couplings and their Witten diagrams}
In the ambient framework it is straightforward to parameterise the most general on-shell cubic interaction up to integrations by parts. As explained in the introduction, on-shell the most general cubic interaction involving fields of spins $s_1$-$s_2$-$s_3$ is given by
\begin{equation}\label{ansatz}
    {\cal V}_{s_1,s_2,s_3} = \sum\limits_{n_i} g^{n_1,n_2,n_3}_{s_1,s_2,s_3}I_{s_1,s_2,s_3}^{n_1,n_2,n_3}(\Phi_i),
    \end{equation}
with 
\begin{multline}
    I_{s_1,s_2,s_3}^{n_1,n_2,n_3}(\Phi_i)=\mathcal{Y}_1^{s_1-n_2-n_3}\mathcal{Y}_2^{s_2-n_3-n_1}\mathcal{Y}_3^{s_3-n_1-n_2}\\\times\mathcal{H}_1^{n_1}\mathcal{H}_2^{n_2}\mathcal{H}_3^{n_3}\,\Phi_1(X_1,U_1)\Phi_2(X_2,U_2)\Phi_3(X_3,U_3)\Big|_{X_i=X}\,.\label{innn}
\end{multline}
In the next section we determine the couplings $g^{n_1,n_2,n_3}_{s_1,s_2,s_3}$ in minimal bosonic higher-spin theory by employing the holographic duality. We do so by matching the tree-level three-point Witten diagram generated by the vertex \eqref{ansatz} with the corresponding three-point function \eqref{dual3pt} in the dual free scalar $O\left(N\right)$ vector model. We use the following two ingredients:
\begin{enumerate}
    \item {\bf Boundary-to-bulk propagators and their derivatives} \\ The spin-$s$ boundary-to-bulk propagator \cite{Mikhailov:2002bp} is a linear solution to the bulk wave equation,
\begin{align}
    \left(\Box - m^2_s\right)\Pi_{\Delta,s} = 0, \qquad m^2_s = \Delta\left(\Delta-d\right)-s,
\end{align}
with a boundary delta-function source. In the ambient framework, it can be expressed in the form 
\begin{equation}
    \Pi_{\Delta,s}(X,P|U,Z)=\frac{C_{\Delta,s}}{s!}\,\frac{\left[(-2P\cdot X)(U\cdot Z)+2(U\cdot P)(Z\cdot X)\right]^s}{(-2P\cdot X)^{\Delta+s}}\,,
\end{equation}
with 
\begin{align}
    Z^2 &= 0\,,& P^2&=0\,,& Z\cdot P&=0\,,
\end{align}
which also ensures tracelessness of the propagator. We use the normalisation \cite{Costa:2014kfa}
\begin{equation}
    C_{\Delta,s}=\frac{(\Delta+s-1)\Gamma(\Delta)}{2\pi^{d/2}(\Delta-1)\Gamma(\Delta+1-\tfrac{d}{2})}\,.
\end{equation}
Just like for the CFT results of the previous section, it is instrumental to express the scalar boundary-to-bulk propagator in the Schwinger-parameterised form
\begin{equation}
    \frac{1}{\left(-2X \cdot P\right)^\Delta} = \frac{1}{\Gamma\left(\Delta\right)} \int^{\infty}_0 \frac{dt}{t} t^{\Delta}e^{2t P \cdot X}.\label{schwingprop1}
\end{equation}
In this way the $n$-th ambient derivative of the bulk-to-boundary propagator can be expressed in terms of a scalar propagator of dimension $\Delta+n$:
\begin{align}
&(W\cdot\pl_X)^n\Pi_{\Delta,s}(X,P|U,Z)\\
&\hspace*{1.5cm}=\frac{C_{\Delta,s}}{s!}\sum_{i=0}^s\sum_{\omega=0}^i\binom{s}{i}\binom{i}{\omega} \frac{2^n}{\left(n-\omega+1\right)_{\omega}}\frac{1}{\Gamma(\Delta+i)}\,(U\cdot P)^i(U\cdot Z)^{s-i}\nonumber\\ \nonumber
& \hspace*{3cm}\times (Z\cdot W)^\omega(P\cdot W)^{n-\omega}\left((Z\cdot\pl_P)^{i-\omega}\,\int_0^\infty\frac{dt}{t}\,t^{\Delta+n}e^{2tP\cdot X}\,\right).\label{schwingprop2}
\end{align}
\item {\bf Three-point bulk integrals}\\
 Employing the above Schwinger representations for the boundary-to-bulk propagators, the integral over AdS-space generated by the generic tensor structure \eqref{ansatz} can be reduced to that of a basic scalar cubic interaction 
\begin{align}
  & \int_{\text{AdS}_{d+1}}dX \left( \prod_{i=1}^3\frac{dt_i}{t_i}\,t_i^{\Delta_i}\right)e^{2(t_1 P_1+t_2 P_2+t_3P_3)\cdot X}\\ \nonumber
  & \hspace*{2cm}=\pi^{\frac{d}{2}}\Gamma\left(\frac{ \sum\nolimits^3_{i=1} \Delta_i-d}{2}\right)  \int^{\infty}_0 \prod\limits^{3}_{i=1}\left(\frac{dt_i}{t_i} t^{\Delta_i}_i\right) e^{\left(-t_1t_2 P_{12} - t_1t_3 P_{13}-t_2t_3 P_{23}\right)},
\end{align}
where we defined $P_{ij}=-2P_i\cdot P_j=x_{ij}^2$. The integrations over the $t_i$ are simply integral representations of Gamma functions.
\end{enumerate}
The resulting amplitude of the $s_1$-$s_2$-$s_3$ three-point Witten diagram for the vertex \eqref{innn} for general $n_i$ is lengthy, and we give it explicitly in appendix \ref{3ptapp}. The crucial observation is that the simplest structure with $n_1 = n_2 = n_3 = 0$ is singled out when demanding that we generate the precise combination of boundary structures of the dual free CFT correlator
\begin{multline}\label{resbulk}
  \int_{\text{AdS}_{d+1}} dX\,I_{s_1,s_2,s_3}^{0,0,0}=\,\frac{\pi ^{\frac{3}{2}-d}\,(-1)^{s_1+s_2+s_3}\,2^{-3 d-s_1 -s_2 -s_1 +8}\,\Gamma (d+s_1 +s_2 +s_1 -3)}{(x_{12}^2)^{d/2-1}(x_{23}^2)^{d/2-1}(x_{31}^2)^{d/2-1}}\,\\\times\frac{ \Gamma (d-3+s_1) \Gamma (d-3+s_2) \Gamma (d-3+s_1) }{\Gamma \left(\frac{d-3}{2}+s_1 \right) \Gamma \left(\frac{d-3}{2}+s_2 \right) \Gamma \left(\frac{d-3}{2}+s_1 \right)}\prod_{i=1}^3\left(q_i^{1-\frac{d}{4}}
J_{\frac{d}{2}-2}\left(\sqrt{q_i}\right) \right)
\,{\sf Y}_1^{s_1}{\sf Y}_2^{s_2}{\sf Y}_3^{s_3}\,.
\end{multline}
This identifies the bulk structure that reproduces the free scalar CFT correlator up to an overall coefficient.\\

\noindent A few remarks are in order:
\begin{itemize}
    \item The dependence on the spins $s_i$ is completely factorised but for the overall Gamma function prefactor coming from the AdS integration.
    \item Remarkably, a single ambient structure resums all couplings including the quasi-mininal structures present in the higher spin theory \cite{Boulanger:2008tg,Bekaert:2010hw,Joung:2011ww}. In Appendix B we outline how the lower derivative terms are generated upon translating the result in terms of AdS covariant derivatives. We also provide a recursive solution for the radial reduction.
    \item We have checked that above couplings are gauge invariant and listed in the classification of \cite{Joung:2011ww}. This shows that the holographic reconstruction at this order is compatible with Noether procedure (see Appendix \ref{GaugeInv}).
    \item Although the couplings we obtained are on-shell, in appendix \ref{Appendix:deDonder} we employ the Noether procedure to establish their off-shell completetion in de Donder gauge.
    \item The couplings considered here were shown to induce deformations to the gauge transformations and to the gauge algebra compatible with the relevant higher-spin algebras (see e.g. \cite{Joung:2013nma}). It was however not possible so far to determine the relative coefficients (though some progress has been made in \cite{Fradkin:1986qy,Vasilev:2011xf,Kessel:2015kna,hsalg}).
\end{itemize}

\section{Holographic reconstruction}

It is now straightforward to combine the above bulk results with those for the three-point functions in the $O\left(N\right)$ model in Section \ref{boundary}. This gives the complete holographic reconstruction of the cubic couplings for the minimal bosonic higher-spin theory in AdS$_{d+1}$.\\

\noindent
Normalising the two-point functions of the ${\cal J}_{s_i}$ to one in both the bulk and boundary computations, we obtain the following coupling constants
\begin{align}
   g_{s_1,s_2,s_3}= \frac{1}{\sqrt{N}}\frac{\pi ^{\frac{d-3}{4}}2^{\tfrac{3 d-1+s_1+s_2+s_3}{2}}}{ \Gamma (d+s_1+s_2+s_3-3)}\prod_{i=1}^3\sqrt{\frac{\Gamma(s_i+\tfrac{d-1}{2})}{\Gamma\left(s_i+1\right)}}\,.
\end{align}
The complete bulk cubic coupling thus reads:\footnote{Analytically continuing our result to odd-spins and introducing internal generators amounts to a further factor of $i^{s_1+s_2+s_3}$ as a consequence of the reality conditions.}
\begin{equation}
    \mathcal{V}=\sum_{s_1,s_2,s_3} g_{s_1,s_2,s_3}I_{s_1,s_2,s_3}^{0,0,0}\,.
\end{equation}
The simplest form for the above coupling manifests itself in AdS$_4$, where the spin-dependence remarkably coincides with the one obtained in \cite{Metsaev:1991mt,Metsaev:1991nb} from a flat space quartic analysis:\footnote{With the appearance of \cite{Bekaert:2015tva} and drawing from the result of \cite{Metsaev:1991mt,Metsaev:1991nb}, for the AdS$_4$ case an educated guess for the relevant part of the spin-dependence was made in \cite{Skvortsov:2015pea}.}
\begin{align}
   g_{s_1,s_2,s_3}= \frac{2^{\tfrac{s_1+s_2+s_3}{2}+4}}{\sqrt{N}\,\Gamma(s_1+s_2+s_3)}\,.
\end{align}
This is in accordance with the flat limit of the above AdS$_4$ vertices. Notice that the flat limit requires to first fix the spin of the external states and keep the highest derivative term for each triple of spins. Furthermore, it is a nice consistency check that the above result reduces to the $0$-$0$-$s$ coupling obtained in \cite{Bekaert:2015tva} for any two pairs of spins $s_i$ set to zero.

\subsection*{Acknowledgements}

C. S. would like to thank X.~Bekaert, J.~Erdmenger and D.~Ponomarev for useful discussions. M.T. is indebted to E.~Joung for useful discussions and comments. The research of M. T. is partially supported by the Fund for Scientific Research-FNRS Belgium, grant FC 6369 and by the Russian Science Foundation grant 14-42-00047 in association with Lebedev Physical Institute. 

\appendix
\section{Generic three-point amplitude}
\label{3ptapp}
The final result for the integral over AdS$_{d+1}$ space of the generic tensor structure $I_{s_1,s_2,s_3}^{n_1,n_2,n_3}$ can be given in the form below:

{\footnotesize
\begin{align}
    \int_{\text{AdS}_{d+1}} dX I_{s_1,s_2,s_3}^{n_1,n_2,n_3}(\Pi_i)&=\mathcal{E}_3\left(\prod_{\alpha=1}^3\sum_{i_\alpha=0}^{s_\alpha-k_\alpha}\sum_{j_\alpha=0}^{k_{\alpha}}\sum_{\delta_{\alpha}=0}^{n_{\alpha}}\sum_{\omega_{\alpha}=0}^{i_{\alpha-1}+j_{\alpha-1}}\right)\sum_{\gamma_3=0}^{i_2+j_2-\omega_3}\sum_{\gamma_1=0}^{i_3+j_3-\omega_1}\sum_{\gamma_2=0}^{i_1+j_1-\omega_2-\gamma_3}\\\nonumber
    &\hspace{-80pt} \frac{n_1!n_2!n_3! (i_1 +j_1)! (i_2 +j_2)! (i_3 +j_3)!(s_3-n_1-n_2)! (s_2-n_1-n_3)! (s_1-n_2 -n_3 )! }{\gamma_1!\gamma_2!\gamma_3!\delta_1!\delta_2!\delta_3!i_1!i_2!i_3!\omega_1!\omega_2!\omega_3}\\\nonumber &\hspace{-80pt}\frac{\Gamma \left(i_1 +j_1 -n_2 +s_1 -\gamma_2  +\delta_{12}-\omega_2  \right) \Gamma \left(i_2 +j_2 -n_3 +s_2 -\gamma_3  +\delta_{23}-\omega_3  \right) \Gamma \left(i_3 +j_3 -n_1+s_3 -\gamma_1  +\delta_{31}-\omega_1  \right)}{\Gamma (j_1 -n_2 +\delta_2  +1) \Gamma (j_2 -n_3 +\delta_3  +1) \Gamma (j_3 -n_1+\delta_1  +1)}\\\nonumber
    &\hspace{-80pt}\frac1{ \Gamma (i_1 +j_1 -\gamma_2  -\gamma_3  -\omega_2  +1)  \Gamma (i_2 +j_2 -\gamma_1  -\gamma_3  -\omega_3  +1)  \Gamma (i_3 +j_3 -\gamma_1  -\gamma_2  -\omega_1  +1)  }\\\nonumber
    &\hspace{-80pt}\frac1{\Gamma (-i_1 -n_2 -n_3 +s_1 -\omega_3  +1)\Gamma (-i_2 -n_1-n_3 +s_2 -\omega_1  +1)\Gamma (-i_3 -n_1-n_2 +s_3 -\omega_2  +1)}
    \\\nonumber
    &\hspace{-80pt}\frac1{\Gamma (-j_1 +n_2 +n_3 -\delta_2  -\delta_3  +1) \Gamma (-j_2 +n_1+n_3 -\delta_1  -\delta_3  +1) \Gamma (-j_3 +n_1+n_2 -\delta_1  -\delta_2  +1)}\\\nonumber
    &\hspace{-80pt}\frac1{ \Gamma (i_1 +j_1 +\Delta_1 ) \Gamma (i_2 +j_2 +\Delta_2 ) \Gamma (i_3 +j_3  +\Delta_3 )}\\\nonumber
    &\hspace{-80pt}(-1)^{-\delta_1 -\delta_2 -\delta_3 +i_1+i_2 +i_3 +j_1 +j_2 +j_3 +n_1 +n_2 +n_3 +s_1 +s_2 +s_3 } 2^{-\gamma_1 -\gamma_2 -\gamma_3 -\delta_1 -\delta_2 -\delta_3 -n_1 -n_2 -n_3 +s_1 +s_2 +s_3 -\omega_1 -\omega_2 -\omega_3 }\\
    &\hspace{-80pt}{\sf H}_1 ^{\gamma_1 +\delta_1 +\omega_1 } {\sf H}_2 ^{\gamma_2 +\delta_2 +\omega_2 } {\sf H}_3 ^{\gamma_3 +\delta_3 +\omega_3 } {\sf Y}_1 ^{s_1-\gamma_2 -\gamma_3 -\delta_2 -\delta_3 -\omega_2 -\omega_3 } {\sf Y}_2 ^{s_2-\gamma_1 -\gamma_3 -\delta_1 -\delta_3 -\omega_1 -\omega_3 } {\sf Y}_3 ^{s_3-\gamma_1 -\gamma_2 -\delta_1 -\delta_2 -\omega_1 -\omega_2 }\,,\nonumber
\end{align}}

\noindent with the prefactor
\begin{multline}
    \mathcal{E}_3=\frac{\pi ^{-d} \Gamma ( \Delta_1-1) \Gamma ( \Delta_2-1) \Gamma ( \Delta_3-1) ( \Delta_1+s_1 -1) ( \Delta_2+s_2 -1) ( \Delta_3+s_3 -1) }{16\, \Gamma \left(\Delta_1+1-\frac{d}{2}\right) \Gamma \left( \Delta_2+1-\frac{d}{2}\right) \Gamma \left(\Delta_3+1-\frac{d}{2}\right) }\\\times\Gamma \left(\frac{\tau_1+ \tau_2+ \tau_3-d}{2} +s_1 +s_2 +s_3 - n_1 - n_2 - n_3 \right)\,\frac1{(x_{12}^2)^{\delta_{12}}(x_{23}^2)^{\delta_{23}}(x_{31}^2)^{\delta_{31}}}\nonumber
\end{multline}
and
\begin{align}
    \delta_{ij}&=\frac12(\tau_i+\tau_j-\tau_k)\,,& \tau_i&=\Delta_i-s_i\,.
\end{align}
\noindent The above result shows how each individual term in the bulk decomposes into conformal structures  upon performing the integral over AdS$_{d+1}$.

\section{Radial Reduction}
\label{radialred}

In this appendix we review the recipe to perform the radial reduction of any ambient vertex of the type used in this letter. We follow the original works \cite{Joung:2011ww,Taronna:2012gb}.
% First of all it is convenient to introduce AdS$_{d+1}$ auxiliary variables which commute with the AdS intrinsic covariant derivative. This can be achieved working with:
% \begin{align}\label{u}
%     u_i^\mu&=u_i^a \,e_{a}{}^\mu(x_i)\,,& \pl_{u_i^\mu}&=\pl_{u^a_i} e^{a}{}_\mu(x_i)\,,
% \end{align}
% in terms of the AdS$_{d+1}$ vielbein $e^{a}{}_\mu(x)$ and the tangent space constant variables $u^a_i$'s. Thanks to the vielbein postulate one can then explicitly verify that
% \begin{align}
%     [\nabla_\mu,u^\nu]&=0\,,& [\nabla_\mu,\pl_{u^\nu}]&=0\,.
% \end{align}
We first introduce the AdS$_{d+1}$ covariant derivative $\nabla_\mu$, acting on functions of the intrinsic auxiliary variable $u^a$ as:
\begin{equation}
    \nabla_\mu=\bar{\nabla}_\mu+\omega_\mu^{ab}\, u_a\,\tfrac{\partial}{\partial u^b}\,,
\end{equation}
with $\omega_\mu^{ab}$ the AdS$_{d+1}$ spin-connection and $\bar{\nabla}_\mu$ the standard covariant derivative acting on tensor indices as usual.\footnote{In more details we recall that:
\begin{align}
    \nabla_\mu f_\nu&=\pl_\mu f_\nu-\Gamma^\rho_{\mu\nu}f_\rho\,,& \Gamma^\rho_{\mu\nu}&=-\,\pl^\rho \hat{X}^M\,\pl_{\mu}\pl_\nu\hat{X}_M\,,& \hat{X}^M=\frac{X^M}{\sqrt{-X^2}}\,.
\end{align}}
One can then arrive to the following dictionary for the radial reduction of ambient operators:
\begin{align}\label{commCov}
    \hat{\pl}^M_{U}&=\pl^M_U-\frac{X^M}{X^2}\,X\cdot\pl_U\,,\\
    \nabla^M&=\pl_X^M-\frac{1}{X^2}(X^M X\cdot\pl_X+U^M X\cdot\pl_U-U\cdot X\pl_U^M)\,,
\end{align}
which satisfy the following relations:
\begin{align}
    [X\cdot\pl_U,\nabla_M]&=0\,,& [\pl_U\cdot\pl_U,\nabla_M]&=0\,,&  [\nabla_M,X^2]&=0\,,&
    X\cdot\nabla&=0\,,
\end{align}
together with
\begin{align}
    [\hat{\pl}_{U}^M,\nabla^N]=\tfrac{X^M}{X^2}\,\hat{\pl}_U^N\,,
\end{align}
While the latter formalism allows us easily to achieve the radial reduction  (since we can keep working directly in the ambient space language), the price to pay is some ordering ambiguities associated to the fact that the tangent intrinsic auxiliary variable $u^a$ does not commute with $\nabla_\mu$ contrary to $u^\mu(x)=u^a e_{a}{}^\mu(x)$. Fixing the ordering ambiguity as:
\begin{multline}
    I_{l_1,l_2,l_3,k_1,k_2,k_3}^{n_1,n_2,n_3,m_1,m_2,m_3}\equiv {\cal H}_1^{n_1} {\cal H}_2^{n_2} {\cal H}_3^{n_3}\,{\cal Y}^{l_1}_1{\cal Y}^{l_2}_2{\cal Y}^{l_3}_3\,\widetilde{\cal Y}^{k_1}_1\widetilde{\cal Y}^{k_2}_2\widetilde{\cal Y}^{k_3}_3\\\times\,(X_1^2)^{-m_1}(X_2^2)^{-m_2}(X_3^2)^{-m_3}\,\Phi_1\,\Phi_2\,\Phi_3\,,
\end{multline}
with
\begin{align}
    \widetilde{\cal Y}_1&=\pl_{U_1}\cdot\nabla_2\,,& \widetilde{\cal Y}_2&=\pl_{U_2}\cdot\nabla_3\,,&\widetilde{\cal Y}_3&=\pl_{U_3}\cdot\nabla_1\,,
\end{align}
one establishes the following recursion relations which iteratively accomplish the radial reduction of the vertices:\footnote{Notice that after replacing all ambient derivatives one can also replace all ambient contractions $\pl_U$ with $\hat{\pl}_U$ for free, owing to \eqref{commCov}.}
{\allowdisplaybreaks
\begin{subequations}
\begin{align}
    I_{l_1,l_2,l_3,0,0,k_3}^{n_1,n_2,n_3,m_1,0,0}&\!= I_{l_1,l_2,l_3-1,0,0,k_3+1}^{n_1,n_2,n_3,m_1,0,0}\!-l_2(\Delta_1\!-\!k_3\!-\!n_2\!-\!2m_1)\,I_{l_1,l_2-1,l_3-1,0,0,k_3}^{n_1+1,n_2,n_3,m_1+1,0,0}\\&-\!n_3(l_3\!-\!1)\,I_{l_1,l_2,l_3-2,0,0,k_3}^{n_1+1,n_2+1,n_3-1,m_1+1,0,0}\nonumber\\\nonumber
    &-l_1(\Delta_2\!-\!l_1\!-\!n_3\!-\!l_3\!+\!2)\,I_{l_1-1,l_2,l_3-1,0,0,k_3}^{n_1,n_2+1,n_3,m_1+1,0,0}\\
    &+\!\lambda\,l_1  l_2(l_3-1)\,I_{l_1-1,l_2-1,l_3-2,0,0,k_3}^{n_1+1,n_2+1,n_3,m_1+1,0,0}\,,\nonumber\\
    I_{l_1,l_2,0,0,k_2,k_3}^{n_1,n_2,n_3,m_1,0,m_3}&\!= I_{l_1,l_2-1,0,0,k_2+1,k_3}^{n_1,n_2,n_3,m_1,0,m_3}\!\!-l_1(\Delta_3\!-\!k_2\!-\!n_1\!-\!2m_3)\,I_{l_1-1,l_2-1,l_3,0,k_2,k_3}^{n_1,n_2,n_3+1,m_1,0,m_3+1}\nonumber\\&-n_2(l_2\!-\!1)\,I_{l_1,l_2-2,l_3,0,k_2,k_3}^{n_1\!+\!1,n_2-1,n_3\!+\!1,m_1,0,m_3\!+\!1}\\
    I_{l_1,0,0,k_1,k_2,k_3}^{n_1,n_2,n_3,m_1,m_2,m_3}&\!= I_{l_1-1,0,0,k_1+1,k_2,k_3}^{n_1,n_2,n_3,m_1,m_2,m_3}-n_1(l_1\!-\!1)\,I_{l_1-2,0,0,k_1,k_2,k_3}^{n_1-1,n_2+1,n_3+1,m_1,m_2+1,m_3}\,.
\end{align}
\end{subequations}}

\noindent
Above $\Delta_i$ are the homogeneity degrees of the various fields and $\lambda$ is an auxiliary variable to be replaced at the very end of the recursion procedure as follows:
\begin{equation}\label{lambda}
    \lambda^n\equiv(-1)^n(\Delta+d)(\Delta+d-2)\ldots(\Delta+d-2n+2)\,,
\end{equation}
where $\Delta$ is the total degree of homogeneity of the given term of the vertex. The reason $\lambda$ appears is that in order to reduce the vertex to the basis chosen one has to perform integration by parts. The above recursion relations solve the problem of reducing a generic ambient vertex in the intrinsic AdS$_{d+1}$ basis $I_{0,0,0,k_1,k_2,k_3}^{n_1,n_2,n_3,0,0,0}$.

Some examples giving the radial reduction of the HS couplings found in this letter can be easily given by implementing the above recursion relations for massless fields ($\Delta_i=2-d-s_i$) into a computer program:
{\allowdisplaybreaks
\begin{subequations}
\begin{align}
    {\cal Y}_1{\cal Y}_2&=\widetilde{\cal Y}_1\widetilde{\cal Y}_2-(d-2) {\cal H}_3\,,\\
    {\cal Y}_1{\cal Y}_2{\cal Y}_3&=\widetilde{\cal Y}_1\widetilde{\cal Y}_2\widetilde{\cal Y}_3-(d-1)({\cal H}_1\widetilde{\cal Y}_1+{\cal H}_2\widetilde{\cal Y}_2+{\cal H}_3\widetilde{\cal Y}_3)\,,\nonumber\\
    {\cal Y}^2_1{\cal Y}^2_2{\cal Y}^2_3
    &=\widetilde{\cal Y}_1^2 \widetilde{\cal Y}_2^2 \widetilde{\cal Y}_3^2-2(2 d+1) {\cal H}_1  \widetilde{\cal Y}_1 ^2 \widetilde{\cal Y}_2  \widetilde{\cal Y}_3 -2(2 d+3) {\cal H}_2  \widetilde{\cal Y}_1  \widetilde{\cal Y}_2 ^2 \widetilde{\cal Y}_3 -2(2 d+1) {\cal H}_3  \widetilde{\cal Y}_1  \widetilde{\cal Y}_2  \widetilde{\cal Y}_3 ^2\nonumber\\
    &+2 d (d+2) {\cal H}_1^2 \widetilde{\cal Y}_1^2+2 d (d+2) {\cal H}_3^2 \widetilde{\cal Y}_3^2+2 d (d+2) {\cal H}_2^2 \widetilde{\cal Y}_2^2\nonumber\\
    &+8 \left(d^2+2 d+2\right) {\cal H}_1  {\cal H}_2  \widetilde{\cal Y}_1  \widetilde{\cal Y}_2 +2\left(4 d^2+4 d-1\right) {\cal H}_1  {\cal H}_3  \widetilde{\cal Y}_1  \widetilde{\cal Y}_3 \nonumber\\
    &+4 (d+1) (2 d+1) {\cal H}_2  {\cal H}_3  \widetilde{\cal Y}_2  \widetilde{\cal Y}_3 -8 \left(d^3+2 d^2+d+1\right) {\cal H}_1  {\cal H}_2  {\cal H}_3\,,\\
    {\cal Y}^3_1{\cal Y}^3_2&{\cal Y}^3_3=\widetilde{\cal Y}_1^3 \widetilde{\cal Y}_2^3 \widetilde{\cal Y}_3^3\nonumber\\
    &-9 (d+2) {\cal H}_1 \widetilde{\cal Y}_1^3 \widetilde{\cal Y}_2^2 \widetilde{\cal Y}_3^2-9 (d+4) {\cal H}_2 \widetilde{\cal Y}_1^2 \widetilde{\cal Y}_2^3 \widetilde{\cal Y}_3^2\nonumber\\
    &-9 (d+2) {\cal H}_3 \widetilde{\cal Y}_1^2 \widetilde{\cal Y}_2^2 \widetilde{\cal Y}_3^3+18 \left(d^2+5 d+5\right) {\cal H}_1^2 \widetilde{\cal Y}_1^3 \widetilde{\cal Y}_2 \widetilde{\cal Y}_3\nonumber\\
    &+27 \left(2 d^2+12 d+21\right) {\cal H}_1 {\cal H}_2 \widetilde{\cal Y}_1^2 \widetilde{\cal Y}_2^2 \widetilde{\cal Y}_3\nonumber\\
    &+18 \left(d^2+7 d+11\right) {\cal H}_2^2 \widetilde{\cal Y}_1 \widetilde{\cal Y}_2^3 \widetilde{\cal Y}_3+18 \left(d^2+5 d+5\right) {\cal H}_3^2 \widetilde{\cal Y}_1 \widetilde{\cal Y}_2 \widetilde{\cal Y}_3^3\nonumber\\
    &+54 (d+1) (d+3) {\cal H}_1 {\cal H}_3 \widetilde{\cal Y}_1^2 \widetilde{\cal Y}_2 \widetilde{\cal Y}_3^2+27 (d+2) (2 d+7) {\cal H}_2 {\cal H}_3 \widetilde{\cal Y}_1 \widetilde{\cal Y}_2^2 \widetilde{\cal Y}_3^2\nonumber\\
    &-54 (d+3) \left(d^2+6 d+10\right) {\cal H}_1^2 {\cal H}_2 \widetilde{\cal Y}_1^2 \widetilde{\cal Y}_2-54 (d+3) \left(d^2+6 d+10\right) {\cal H}_1 {\cal H}_2^2 \widetilde{\cal Y}_1 \widetilde{\cal Y}_2^2\nonumber\\
    &-54 (2 d+7) \left(2 d^2+8 d+9\right) {\cal H}_1 {\cal H}_2 {\cal H}_3 \widetilde{\cal Y}_1 \widetilde{\cal Y}_2 \widetilde{\cal Y}_3-54 (d+4) \left(d^2+3 d+1\right) {\cal H}_1 {\cal H}_3^2 \widetilde{\cal Y}_1 \widetilde{\cal Y}_3^2\nonumber\\
    &-27 \left(2 d^3+14 d^2+28 d+11\right) {\cal H}_1^2 {\cal H}_3 \widetilde{\cal Y}_1^2 \widetilde{\cal Y}_3-27 \left(2 d^3+16 d^2+38 d+25\right) {\cal H}_2^2 {\cal H}_3 \widetilde{\cal Y}_2^2 \widetilde{\cal Y}_3\nonumber\\
    &-54 \left(d^3+8 d^2+21 d+17\right) {\cal H}_2 {\cal H}_3^2 \widetilde{\cal Y}_2 \widetilde{\cal Y}_3^2-6 (d+1) (d+3) (d+5) {\cal H}_1^3 \widetilde{\cal Y}_1^3\nonumber\\
    &-6 (d+1) (d+3) (d+5) {\cal H}_2^3 \widetilde{\cal Y}_2^3-6 (d+1) (d+3) (d+5) {\cal H}_3^3 \widetilde{\cal Y}_3^3\nonumber\\
    &+54 (d+3) \left(2 d^3+15 d^2+34 d+27\right) {\cal H}_1^2 {\cal H}_2 {\cal H}_3 \widetilde{\cal Y}_1\nonumber\\
    &+54 (d+3) \left(2 d^3+14 d^2+28 d+19\right) {\cal H}_1 {\cal H}_2^2 {\cal H}_3 \widetilde{\cal Y}_2\nonumber\\
    &+54 (d+3) \left(2 d^3+14 d^2+29 d+19\right) {\cal H}_1 {\cal H}_2 {\cal H}_3^2 \widetilde{\cal Y}_3\,.
\end{align}
\end{subequations}}

\noindent
Notice that the result is not manifestly cyclic as a consequence of the ordering ambiguity of the derivatives $\pl_U$. This complicates the computation in the generic case. On the other hand the above structures are manifestly intrinsic expressions and we have carefully checked their gauge invariance (see e.g. Appendix \ref{GaugeInv}).

\section{Gauge Invariance and the Vertices Classification}\label{GaugeInv}
It is straightforward to relate the result obtained in this letter to the previous classification of cubic gauge-invariant couplings given in\footnote{Notice the change of notation $Z_i\rightarrow {\cal H}_i$ to avoid confusion with the ambient boundary auxiliary variables $Z_i$.} \cite{Joung:2011ww}. The change of basis must take into account the fact that the degree of homogeneity was chosen to be
\begin{equation}
    \Delta_s=s-2\,.
\end{equation}
This choice simplified the analysis of the gauge invariance.
In the classification of \cite{Joung:2011ww} one then arrives to the following list of couplings:
\begin{subequations}\label{oldB}
\begin{align}
	C&=e^{\lambda\mathfrak{D}}K\big(\, {\cal Y}_{i}\,,\,{\cal G}\,\big)\Big|_{{\cal G}={\cal H}_1{\cal Y}_1+{\cal H}_2{\cal Y}_2+{\cal H}_3{\cal Y}_3}\,,
	\label{amb K}\\
	\mathfrak{D}&=\left({\cal H}_1\pl_{{\cal Y}_2}\pl_{{\cal Y}_3}+{\cal H}_2{\cal H}_3\pl_{{\cal Y}_1}\pl_{{\cal G}}+\text{cycl.}\right)+{\cal H}_1{\cal H}_2{\cal H}_3\pl_{{\cal G}}^2\,,
\end{align}
\end{subequations}
where ${\cal G}$ is given by
\begin{align}
 	{\cal G} &=
	{\cal Y}_{1}\,{\cal H}_{1} +
	{\cal Y}_{2}\,{\cal H}_{2} +
	{\cal Y}_{3}\,{\cal H}_{3}\,,
	\label{amb G}
\end{align}
Above, ${\cal D}$ is an operator generating a lower derivative tail and $\lambda$ is defined in \eqref{lambda}.
The $K$ is an arbitrary function not fixed by gauge invariance and encoding the relative coefficient of the $s_3+1$ ($s1\geq s_2\geq s_3$) gauge invariant couplings for each triplet $(s_1,s_2,s_3)$ of spins.

It is now lengthy but straightforward to rewrite in the above basis \eqref{oldB} the vertices we have obtained in this letter. The end result is quite simple and shows how the holographically reconstructed vertices \eqref{vert} are indeed gauge-invariant and compatible with the bulk Noether procedure:
\begin{align}
\left({\cal Y}_1^{s_1}{\cal Y}_2^{s_2}{\cal Y}_3^{s_3}\right)_{\Delta_i=d-2-s_i}\qquad\longrightarrow\qquad K_{s_1,s_2,s_3}(Y,G)=e^{-2\,{\cal G}\,\pl_{{\cal Y}_1}\pl_{{\cal Y}_3}\pl_{{\cal Y}_3}}\ {\cal Y}_1^{s_1}{\cal Y}_2^{s_2}{\cal Y}_3^{s_3}\,.
\end{align}
Few remarks are in order:
\begin{itemize}
\item Both for $1$-$1$-$1$ and $2$-$2$-$2$ cases the Yang-Mills ${\cal G}$ and Einstein-Hilbert ${\cal G}^2$ terms are recovered respectively, supplemented by other higher-derivative terms.
\item For any triplet of spins, the quasi-minimal coupling with at most $s_1+s_2-s_3$ ($s_1\geq s_2\geq s_3$) derivatives is always generated.
\item In the basis \eqref{oldB} all the $d$ dependence is reabsorbed into $\lambda$ and the relative coefficients of the different structures are $d$-independent and resummed by a simple exponential.
\end{itemize}

\section{HS theory couplings in the de Donder gauge}\label{Appendix:deDonder}

In this appendix we consider the completion of the couplings found in this paper to the de Donder gauge. This is sufficient for perturbative computations at tree level beyond the cubic order. We work in the formalism of \cite{Joung:2011ww,Joung:2012fv,Taronna:2012gb,Joung:2013nma}, which allows a convenient treatment of gauge invariance and can be easily extended to the de Donder gauge. To this end it is sufficient to recall the ambient space form of the Fronsdal and de Donder operators:
\begin{align}
    \mathcal{F}(\Phi)&=\left[\Box-U\cdot\pl_X\, \mathcal{D}\right]\Phi(X,U)\,,& \mathcal{D}&=\pl_U\cdot\pl_X-\frac12\, U\cdot\pl_X\,\pl_U^2\,.
\end{align}
Furthermore, in the de Donder gauge the leftover gauge transformations:
\begin{align}
    \delta\Phi(X,U)&=U\cdot\pl_X\,E(X,U)\,,&\delta \Phi^\prime(X,U)&=2\,\pl_U\cdot\pl_X\, E(X,U)\,,& (\Phi^\prime&\equiv \pl_U^2 \Phi)\,.
\end{align}
are constrained by $\Box E(X,U)=0$ on top of the standard tracelessness condition $\pl_U^2 \,E(X,U)=0$ for the gauge parameter. The gauge parameter is otherwise \emph{not} divergenceless and as usual the Fronsdal field is constrained to be double-traceless: $(\pl_U^2)^2\,\Phi=0$. Introducing the point splitting notation:
\begin{align}
    {\cal A}_i&=\pl_{U_i}\cdot\pl_{U_i}\,,& {\cal Q}_i&=\pl_{U_i}\cdot\pl_{X_i}\,,
\end{align}
for the trace and divergence operators, the most general ansatz for a vertex in the de Donder gauge only involves $\mathcal{A}_i$ linearly since $\mathcal{A}_i^2\sim 0$. Considering a functional representation for the vertex given by $C_{\text{D}}({\cal Y}_i,{\cal H}_i,{\cal A}_i)$, the gauge invariance condition reads in the de Donder gauge as follows:
{\allowdisplaybreaks
\begin{align}
    &\left[\mathcal{Y}_3\pl_{\mathcal{H}_2}-\mathcal{Y}_2\pl_{\mathcal{H}_3}+\lambda(\mathcal{Y}_2\pl_{\mathcal{Y}_2}-\mathcal{Y}_3\pl_{\mathcal{Y}_3}+\mathcal{A}_2\pl_{\mathcal{A}_2}-\mathcal{A}_3\pl_{\mathcal{A}_3})\pl_{\mathcal{Y}_1}\right]C_{\text{D}}\,E_1\Phi_2\Phi_3\nonumber\\
    -&\frac12 \mathcal{A}_2\left[\mathcal{Y}_1\pl_{\mathcal{H}_3}-\mathcal{Y}_3\pl_{\mathcal{H}_1}+\lambda(\mathcal{Y}_3\pl_{\mathcal{Y}_3}-\mathcal{Y}_1\pl_{\mathcal{Y}_1}+\mathcal{A}_3\pl_{\mathcal{A}_3})\pl_{\mathcal{Y}_2}\right]\pl_{\mathcal{H}_3}C_{\text{D}}\,E_1\Phi_2\Phi_3\nonumber\\
    +&\frac12 \mathcal{A}_2\mathcal{A}_3\left[\mathcal{Y}_2\pl_{\mathcal{H}_1}-\mathcal{Y}_1\pl_{\mathcal{H}_2}+\lambda(\mathcal{Y}_1\pl_{\mathcal{Y}_1}-\mathcal{Y}_2\pl_{\mathcal{Y}_2})\pl_{\mathcal{Y}_3}\right]\pl_{\mathcal{H}_3}\pl_{\mathcal{H}_1}C_{\text{D}}\,E_1\Phi_2\Phi_3\nonumber\\
    +&2\,\mathcal{Q}_1\left(\pl_{\mathcal{A}_1}-\frac18 \mathcal{A}_2\mathcal{A}_3\,\pl_{\mathcal{H}_1}\pl_{\mathcal{H}_2}\pl_{\mathcal{H}_3}\right)C_{\text{D}}\,E_1\Phi_2\Phi_3=0\,.\label{deDonder}
\end{align}}

\noindent
Above $\lambda$ is defined as in \eqref{lambda}. Given a transverse and traceless vertex $C_{\text{TT}}$, which by definition satisfies
\begin{equation}
    \left[\mathcal{Y}_3\pl_{\mathcal{H}_2}-\mathcal{Y}_2\pl_{\mathcal{H}_3}+\lambda(\mathcal{Y}_2\pl_{\mathcal{Y}_2}-\mathcal{Y}_3\pl_{\mathcal{Y}_3})\pl_{\mathcal{Y}_1}\right]C_{\text{TT}}=0\,,
\end{equation}
together with their cyclic analogues, one can check that the general solution to \eqref{deDonder} is given by:
\begin{equation}
    C_{\text{D}}=\left(1+\frac18\,\mathcal{A}_1\,\mathcal{A}_2\,\mathcal{A}_3\,\pl_{\mathcal{H}_1}\,\pl_{\mathcal{H}_2}\,\pl_{\mathcal{H}_3}\right)C_{\text{TT}}\,,
\end{equation}
for any gauge invariant transverse and traceless coupling $C_{\text{TT}}$.
Furthermore, in the flat limit ($\lambda=0$), the above nicely matches the off-shell solution in \cite{Manvelyan:2010jr,Sagnotti:2010at} gauge-fixed to the de Donder gauge.

To summarise, plugging into the above formula the vertices we get from the holographic reconstruction, we obtain the following de Donder gauge vertices for the minimal HS theory under consideration:
\begin{subequations}
\begin{align}
    (C_{\text{D}})_{s_1,s_2,s_3}&=\left(1+\frac18\,\mathcal{A}_1\,\mathcal{A}_2\,\mathcal{A}_3\,\pl_{\mathcal{H}_1}\,\pl_{\mathcal{H}_2}\,\pl_{\mathcal{H}_3}\right)e^{\widetilde{\mathfrak{D}}}\,\mathcal{Y}_1^{s_1}\mathcal{Y}_1^{s_2}\mathcal{Y}_1^{s_3}\,,\\
    \widetilde{\mathfrak{D}}&=\lambda\left(\mathcal{H}_1\pl_{\mathcal{Y}_2}\pl_{\mathcal{Y}_3}-2\,\mathcal{H}_2\mathcal{H}_3\pl_{\mathcal{Y}_2}\pl_{\mathcal{Y}_3}\pl_{\mathcal{Y}_1}^2+\text{cycl.}\right)\\&\hspace{100pt}+4\,\lambda\mathcal{H}_1\mathcal{H}_2\mathcal{H}_3\,\pl_{\mathcal{Y}_1}^2\pl_{\mathcal{Y}_2}^2\pl_{\mathcal{Y}_3}^2-2\,\mathcal{G}\,\pl_{\mathcal{Y}_1}\pl_{\mathcal{Y}_2}\pl_{\mathcal{Y}_3}\,.\nonumber
\end{align}
\end{subequations}
Nicley, the vertex involving the trace components of the Fronsdal fields is obtained from the TT vertex through the action of a very simple differential operator.
\iffalse
Remarkably, the structure of the cubic vertex is similar as at the quadratic order where:
\begin{equation}
    \mathcal{L}_s^{(2)}=\frac{s!}2\,\left(\Phi(X,\pl_U)\,\Box\,\Phi(X,U)-\frac14\,\Phi^\prime(X,\pl_U)\,\Box\,\Phi^\prime(X,U) \right)\,.
\end{equation}
Nicely this feature holds both on flat and AdS background while the vertex involving the trace components of the Fronsdal fields is obtained from the TT vertex through the action of a very simple differential operator.
\fi
\bibliography{refs}

\providecommand{\href}[2]{#2}\begingroup\raggedright\begin{thebibliography}{10}

\bibitem{Vasiliev:1990en}
M.~A. Vasiliev, {\it {Consistent equation for interacting gauge fields of all
  spins in (3+1)-dimensions}},  {\em Phys. Lett.} {\bf B243} (1990) 378--382.

\bibitem{Fradkin:1986qy}
E.~S. Fradkin and M.~A. Vasiliev, {\it {Cubic Interaction in Extended Theories
  of Massless Higher Spin Fields}},  {\em Nucl. Phys.} {\bf B291} (1987) 141.

\bibitem{Vasilev:2011xf}
M.~A. Vasiliev, {\it {Cubic Vertices for Symmetric Higher-Spin Gauge Fields in
  $(A)dS_d$}},  {\em Nucl. Phys.} {\bf B862} (2012) 341--408,
  [\href{http://xxx.lanl.gov/abs/1108.5921}{{\tt arXiv:1108.5921}}].

\bibitem{Boulanger:2013zza}
N.~Boulanger, D.~Ponomarev, E.~D. Skvortsov, and M.~Taronna, {\it {On the
  uniqueness of higher-spin symmetries in AdS and CFT}},  {\em Int. J. Mod.
  Phys.} {\bf A28} (2013) 1350162,
  [\href{http://xxx.lanl.gov/abs/1305.5180}{{\tt arXiv:1305.5180}}].

\bibitem{Kessel:2015kna}
P.~Kessel, G.~Lucena~G\'omez, E.~Skvortsov, and M.~Taronna, {\it {Higher Spins
  and Matter Interacting in Dimension Three}},  {\em JHEP} {\bf 11} (2015) 104,
  [\href{http://xxx.lanl.gov/abs/1505.0588}{{\tt arXiv:1505.0588}}].

\bibitem{hsalg}
C.~Sleight and M.~Taronna, {\it {Scalar Back-reaction from Higher-Spin
  Alegbras: Unfolding and the Cubic Action}},  {\em To appear}.

\bibitem{Sezgin:2002rt}
E.~Sezgin and P.~Sundell, {\it {Massless higher spins and holography}},  {\em
  Nucl. Phys.} {\bf B644} (2002) 303--370,
  [\href{http://xxx.lanl.gov/abs/hep-th/0205131}{{\tt hep-th/0205131}}].
  [Erratum: Nucl. Phys.B660,403(2003)].

\bibitem{Klebanov:2002ja}
I.~R. Klebanov and A.~M. Polyakov, {\it {AdS dual of the critical O(N) vector
  model}},  {\em Phys. Lett.} {\bf B550} (2002) 213--219,
  [\href{http://xxx.lanl.gov/abs/hep-th/0210114}{{\tt hep-th/0210114}}].

\bibitem{Joung:2011ww}
E.~Joung and M.~Taronna, {\it {Cubic interactions of massless higher spins in
  (A)dS: metric-like approach}},  {\em Nucl. Phys.} {\bf B861} (2012) 145--174,
  [\href{http://xxx.lanl.gov/abs/1110.5918}{{\tt arXiv:1110.5918}}].

\bibitem{Joung:2012fv}
E.~Joung, L.~Lopez, and M.~Taronna, {\it {Solving the Noether procedure for
  cubic interactions of higher spins in (A)dS}},  {\em J. Phys.} {\bf A46}
  (2013) 214020, [\href{http://xxx.lanl.gov/abs/1207.5520}{{\tt
  arXiv:1207.5520}}].

\bibitem{Joung:2013nma}
E.~Joung and M.~Taronna, {\it {Cubic-interaction-induced deformations of
  higher-spin symmetries}},  {\em JHEP} {\bf 03} (2014) 103,
  [\href{http://xxx.lanl.gov/abs/1311.0242}{{\tt arXiv:1311.0242}}].

\bibitem{Petkou:2003zz}
A.~C. Petkou, {\it {Evaluating the AdS dual of the critical O(N) vector
  model}},  {\em JHEP} {\bf 03} (2003) 049,
  [\href{http://xxx.lanl.gov/abs/hep-th/0302063}{{\tt hep-th/0302063}}].

\bibitem{Bekaert:2015tva}
X.~Bekaert, J.~Erdmenger, D.~Ponomarev, and C.~Sleight, {\it {Quartic AdS
  Interactions in Higher-Spin Gravity from Conformal Field Theory}},  {\em
  JHEP} {\bf 11} (2015) 149, [\href{http://xxx.lanl.gov/abs/1508.0429}{{\tt
  arXiv:1508.0429}}].

\bibitem{Skvortsov:2015lja}
E.~D. Skvortsov and M.~Taronna, {\it {On Locality, Holography and Unfolding}},
  {\em JHEP} {\bf 11} (2015) 044,
  [\href{http://xxx.lanl.gov/abs/1508.0476}{{\tt arXiv:1508.0476}}].

\bibitem{Taronna:2016ats}
M.~Taronna, {\it {Pseudo-local Theories: A Functional Class Proposal}},  in
  {\em {International Workshop on Higher Spin Gauge Theories Singapore,
  Singapore, November 4-6, 2015}}, 2016.
\newblock \href{http://xxx.lanl.gov/abs/1602.0856}{{\tt arXiv:1602.0856}}.

\bibitem{Bekaert:2014cea}
X.~Bekaert, J.~Erdmenger, D.~Ponomarev, and C.~Sleight, {\it {Towards
  holographic higher-spin interactions: Four-point functions and higher-spin
  exchange}},  {\em JHEP} {\bf 1503} (2015) 170,
  [\href{http://xxx.lanl.gov/abs/1412.0016}{{\tt arXiv:1412.0016}}].

\bibitem{Bekaert:2016ezc}
X.~Bekaert, J.~Erdmenger, D.~Ponomarev, and C.~Sleight, {\it {Bulk quartic
  vertices from boundary four-point correlators}},  in {\em {International
  Workshop on Higher Spin Gauge Theories Singapore, Singapore, November 4-6,
  2015}}, 2016.
\newblock \href{http://xxx.lanl.gov/abs/1602.0857}{{\tt arXiv:1602.0857}}.

\bibitem{Skvortsov:2015pea}
E.~D. Skvortsov, {\it {On (Un)Broken Higher-Spin Symmetry in Vector Models}},
  \href{http://xxx.lanl.gov/abs/1512.0599}{{\tt arXiv:1512.0599}}.

\bibitem{Leigh:2003gk}
R.~G. Leigh and A.~C. Petkou, {\it {Holography of the N=1 higher spin theory on
  AdS(4)}},  {\em JHEP} {\bf 06} (2003) 011,
  [\href{http://xxx.lanl.gov/abs/hep-th/0304217}{{\tt hep-th/0304217}}].

\bibitem{Sezgin:2003pt}
E.~Sezgin and P.~Sundell, {\it {Holography in 4D (super) higher spin theories
  and a test via cubic scalar couplings}},  {\em JHEP} {\bf 07} (2005) 044,
  [\href{http://xxx.lanl.gov/abs/hep-th/0305040}{{\tt hep-th/0305040}}].

\bibitem{Diaz:2006nm}
D.~E. Diaz and H.~Dorn, {\it {On the AdS higher spin / O(N) vector model
  correspondence: Degeneracy of the holographic image}},  {\em JHEP} {\bf 07}
  (2006) 022, [\href{http://xxx.lanl.gov/abs/hep-th/0603084}{{\tt
  hep-th/0603084}}].

\bibitem{Osborn:1993cr}
H.~Osborn and A.~C. Petkou, {\it {Implications of conformal invariance in field
  theories for general dimensions}},  {\em Annals Phys.} {\bf 231} (1994)
  311--362, [\href{http://xxx.lanl.gov/abs/hep-th/9307010}{{\tt
  hep-th/9307010}}].

\bibitem{Erdmenger:1996yc}
J.~Erdmenger and H.~Osborn, {\it {Conserved currents and the energy momentum
  tensor in conformally invariant theories for general dimensions}},  {\em
  Nucl. Phys.} {\bf B483} (1997) 431--474,
  [\href{http://xxx.lanl.gov/abs/hep-th/9605009}{{\tt hep-th/9605009}}].

\bibitem{Stanev:2012nq}
Y.~S. Stanev, {\it {Correlation Functions of Conserved Currents in Four
  Dimensional Conformal Field Theory}},  {\em Nucl. Phys.} {\bf B865} (2012)
  200--215, [\href{http://xxx.lanl.gov/abs/1206.5639}{{\tt arXiv:1206.5639}}].

\bibitem{Zhiboedov:2012bm}
A.~Zhiboedov, {\it {A note on three-point functions of conserved currents}},
  \href{http://xxx.lanl.gov/abs/1206.6370}{{\tt arXiv:1206.6370}}.

\bibitem{Anselmi:1999bb}
D.~Anselmi, {\it {Higher spin current multiplets in operator product
  expansions}},  {\em Class.Quant.Grav.} {\bf 17} (2000) 1383--1400,
  [\href{http://xxx.lanl.gov/abs/hep-th/9906167}{{\tt hep-th/9906167}}].

\bibitem{Craigie:1983fb}
N.~S. Craigie, V.~K. Dobrev, and I.~T. Todorov, {\it {Conformally Covariant
  Composite Operators in Quantum Chromodynamics}},  {\em Annals Phys.} {\bf
  159} (1985) 411--444.

\bibitem{Giombi:2011rz}
S.~Giombi, S.~Prakash, and X.~Yin, {\it {A Note on CFT Correlators in Three
  Dimensions}},  {\em JHEP} {\bf 07} (2013) 105,
  [\href{http://xxx.lanl.gov/abs/1104.4317}{{\tt arXiv:1104.4317}}].

\bibitem{Colombo:2012jx}
N.~Colombo and P.~Sundell, {\it {Higher Spin Gravity Amplitudes From Zero-form
  Charges}},  \href{http://xxx.lanl.gov/abs/1208.3880}{{\tt arXiv:1208.3880}}.

\bibitem{Didenko:2012tv}
V.~E. Didenko and E.~D. Skvortsov, {\it {Exact higher-spin symmetry in CFT: all
  correlators in unbroken Vasiliev theory}},  {\em JHEP} {\bf 04} (2013) 158,
  [\href{http://xxx.lanl.gov/abs/1210.7963}{{\tt arXiv:1210.7963}}].

\bibitem{Costa:2011mg}
M.~S. Costa, J.~Penedones, D.~Poland, and S.~Rychkov, {\it {Spinning Conformal
  Correlators}},  {\em JHEP} {\bf 11} (2011) 071,
  [\href{http://xxx.lanl.gov/abs/1107.3554}{{\tt arXiv:1107.3554}}].

\bibitem{Penedones:2010ue}
J.~Penedones, {\it {Writing CFT correlation functions as AdS scattering
  amplitudes}},  {\em JHEP} {\bf 03} (2011) 025,
  [\href{http://xxx.lanl.gov/abs/1011.1485}{{\tt arXiv:1011.1485}}].

\bibitem{Paulos:2011ie}
M.~F. Paulos, {\it {Towards Feynman rules for Mellin amplitudes}},  {\em JHEP}
  {\bf 10} (2011) 074, [\href{http://xxx.lanl.gov/abs/1107.1504}{{\tt
  arXiv:1107.1504}}].

\bibitem{Taronna:2012gb}
M.~Taronna, {\em {Higher-Spin Interactions: three-point functions and beyond}}.
\newblock PhD thesis, Pisa, Scuola Normale Superiore, 2012.
\newblock \href{http://xxx.lanl.gov/abs/1209.5755}{{\tt arXiv:1209.5755}}.

\bibitem{Costa:2014kfa}
M.~S. Costa, V.~Gon{\c c}alves, and J.~Penedones, {\it {Spinning AdS
  Propagators}},  {\em JHEP} {\bf 1409} (2014) 064,
  [\href{http://xxx.lanl.gov/abs/1404.5625}{{\tt arXiv:1404.5625}}].

\bibitem{Mikhailov:2002bp}
A.~Mikhailov, {\it {Notes on higher spin symmetries}},
  \href{http://xxx.lanl.gov/abs/hep-th/0201019}{{\tt hep-th/0201019}}.

\bibitem{Boulanger:2008tg}
N.~Boulanger, S.~Leclercq, and P.~Sundell, {\it {On The Uniqueness of Minimal
  Coupling in Higher-Spin Gauge Theory}},  {\em JHEP} {\bf 08} (2008) 056,
  [\href{http://xxx.lanl.gov/abs/0805.2764}{{\tt arXiv:0805.2764}}].

\bibitem{Bekaert:2010hw}
X.~Bekaert, N.~Boulanger, and P.~Sundell, {\it {How higher-spin gravity
  surpasses the spin two barrier: no-go theorems versus yes-go examples}},
  {\em Rev. Mod. Phys.} {\bf 84} (2012) 987--1009,
  [\href{http://xxx.lanl.gov/abs/1007.0435}{{\tt arXiv:1007.0435}}].

\bibitem{Metsaev:1991mt}
R.~R. Metsaev, {\it {Poincare invariant dynamics of massless higher spins:
  Fourth order analysis on mass shell}},  {\em Mod. Phys. Lett.} {\bf A6}
  (1991) 359--367.

\bibitem{Metsaev:1991nb}
R.~R. Metsaev, {\it {S matrix approach to massless higher spins theory. 2: The
  Case of internal symmetry}},  {\em Mod. Phys. Lett.} {\bf A6} (1991)
  2411--2421.

\bibitem{Manvelyan:2010jr}
R.~Manvelyan, K.~Mkrtchyan, and W.~Ruhl, {\it {General trilinear interaction
  for arbitrary even higher spin gauge fields}},  {\em Nucl. Phys.} {\bf B836}
  (2010) 204--221, [\href{http://xxx.lanl.gov/abs/1003.2877}{{\tt
  arXiv:1003.2877}}].

\bibitem{Sagnotti:2010at}
A.~Sagnotti and M.~Taronna, {\it {String Lessons for Higher-Spin
  Interactions}},  {\em Nucl. Phys.} {\bf B842} (2011) 299--361,
  [\href{http://xxx.lanl.gov/abs/1006.5242}{{\tt arXiv:1006.5242}}].

\end{thebibliography}\endgroup
\bibliographystyle{JHEP}

\end{document}